TITLE: The Role of Affective States in Computational Psychiatry

AUTHORS: David Benrimoh[1,2], Ryan Smith[3,4], Andreea O. Diaconescu[5,6], Timothy Friesen[2], Sara Jalali[2], Nace Mikus[7], Laura Gschwandtner[8], Jay Gandhi[9], Guillermo Horga[10], Albert Powers[9]

Affiliations:
1. McGill University, Department of Psychiatry, Montreal, Canada
2. Douglas Research Center, Montreal, Canada
3. Laureate Institute for Brain Research, Tulsa, OK, USA
4. Oxley College of Health and Natural Sciences, University of Tulsa, Tulsa, OK, USA
5. Krembil Centre for Neuroinformatics, CAMH, Toronto, Canada
6. Department of Psychiatry, University of Toronto, Toronto, ON, Canada
7. Interacting Minds Centre, Aarhus University, Aarhus, Denmark
8. Department of Cognition- Emotion- and Methods in Psychology, University of Vienna, Vienna, Austria
9. Yale University School of Medicine, New Haven, CT, USA
10. Department of Psychiatry, Columbia University, New York, NY, USA

COIs:
DB is a shareholder and founder of Aifred Health, a digital mental health company which was not involved in this work. NM is a shareholder and founder of TiliaHealth, a digital mental health company which was not involved in this work.

Funding:
DB is supported by an FRQS Junior 1 Clinician-Researcher grant and is a Sidney R. Baer Young Investigator supported by a Brain and Behavior Research Foundation Young Investigator Award.

ABSTRACT:
Studying psychiatric illness has often been limited by difficulties in connecting symptoms and behavior to neurobiology. Computational psychiatry approaches promise to bridge this gap by providing formal accounts of the latent information processing changes that underlie the development and maintenance of psychiatric phenomena. Models based on these theories generate individual-level parameter estimates which can then be tested for relationships to neurobiology. In this review, we explore computational modelling approaches to one key aspect of health and illness: affect. We discuss strengths and limitations of key approaches to modelling affect, with a focus on reinforcement learning, active inference, the hierarchical gaussian filter, and drift-diffusion models. We find that, in this literature, affect is an important source of modulation in decision making, and has a bidirectional influence on how individuals infer both internal and external states. Highlighting the potential role of affect in information processing changes underlying symptom development, we extend an existing model of psychosis, where affective changes are influenced by increasing cortical noise and consequent increases in either perceived environmental instability or expected noise in sensory input, becoming part of a self-reinforcing process generating negatively valenced, over-weighted priors underlying positive symptom development. We then provide testable predictions from this model at computational, neurobiological, and phenomenological levels of description.

INTRODUCTION:

The study of psychiatric disease has traditionally been split across a number of sub-disciplines, from the social to the molecular[1]. Integrating insights from across these disparate levels of explanation has been challenging, in large part because we have lacked a language that can translate findings from different levels into a common framework[2]. The field of computational psychiatry has sought to provide this framework through well-defined mathematical and algorithmic models that can be fit to data from various domains, such as neural responses, choice behavior, and self-report. Algorithmic processes and free parameters in models capturing such processes can generate individual patient-level insights, and these processes can simultaneously be mapped to the neural systems that implement them. Specifically, free parameters in these models can be estimated at the individual patient level, providing an individual-level measure of a given latent process. Thus, these models can serve to provide formal links between behavioral, neurobiological, phenomenological, and social findings[3]. One persistent challenge, however, has been the adequate representation of core psychiatric phenomena within computational models. In this review, we focus on efforts to model one complex, but critical, element of human experience commonly affected within psychopathology: affect[4].

Affect is often understood as an individual's momentary experience of valence (pleasant/unpleasant) and arousal (high/low). While these basic feelings accompany a wide range of other experiences, they are a core feature of emotions. Theories of emotion vary, but are often associated with specific situational causes, cognitions, bodily/interoceptive responses, and observable behaviors (e.g., facial expressions, approach-avoidance drives). A related construct is mood, which typically refers to more stable, long-term affective states that are influenced by cumulative experience. Here we focus on emotions and mood as phenomena with affect as a core element, with the understanding that affect is a broader category and that there are important differences in timescale. We also consider specific emotion categories, as well as relevant cognitive and interoceptive components[5].

One might argue that a true understanding of any cognitive process, especially in the context of psychiatric illness, can be obtained only when its interaction with affect has been adequately modelled. Indeed, affect is known to influence behavior, thoughts, attention, visceral states, perception, and decision-making [6–8]. Executive function is also so closely tied to affect that it is often studied as either 'cold' (affectively neutral) or 'hot' (affectively charged)[9,10]. For example, negatively valenced states with low arousal (boredom) vs. high arousal (e.g., frustration) have been shown to influence experimental results in important ways[11,12].

Affect is a key feature of many psychiatric illnesses, from depressive disorders to the psychosis prodrome[13,14]. Indeed, the prevalence of disturbances in affect and mood in schizophrenia has been reported to be around 40%[15] and up to 80% of patients with chronic schizophrenia have experienced a depressive episode[16]. Similarly, negative symptoms in schizophrenia are characterized by anticipatory anhedonia[17]. In addition, reducing the negative affect or distress associated with symptoms, beyond reduction of the symptoms themselves, has been proposed to be one of the driving factors of the efficacy of psychotropics[18].

In this review, we will discuss previous approaches to modelling affect in computational psychiatry. We will then discuss limitations of this work, and use these limitations to highlight the potential for further advances. In service of this, we will examine the test case of the psychosis prodrome, demonstrating how the role of affect can be integrated into a recently developed model of psychotic symptom development [2].

*Literature Review of Affect in Computational Psychiatry*

Affect is often assumed to exist because it carries information for an organism that is relevant to survival or reproduction. This information (e.g., regarding the presence or degree of threat, the internal states and intentions of others, or the degree to which we find ourselves in a favorable environmental state), can be crucial for guiding behavior adaptively[19,20]. Yet, it can also guide behavior maladaptively, such as in the context of psychiatric illness. If we consider affect as a carrier of information used to guide inference and choice (among other internal processes), models of the associated computational processes could offer important insights.

In this section, we will explore several approaches to modelling affect. We will begin with *reinforcement learning* (RL) models, which generally focus on appetitive and aversive aspects of affect. In this framework, affect is often conceptualized as a way to modify sensitivity to rewards. We will then discuss *active inference*[21], which further posits that agents aim to reduce their uncertainty and maximize their ability to prospectively predict choice outcomes. This often takes the form of first acting to gain information (exploration) and then using that information to achieve reward and remain within desired states (exploitation). In active inference, it has often been posited that information contained within affect assists in inference over both internal and external states, and in predicting the likely outcomes of actions. Following active inference, we will also discuss a related model widely applied to behavioral and neural data, the *hierarchical gaussian filter,* or HGF[22].

A complementary modelling framework we will consider, drift-diffusion models (DDMs), focus on sequential sampling for evidence accumulation toward a choice threshold. DDMs incorporate parameters reflecting the rate of evidence accumulation (drift rate), the starting bias in evidence accumulation, and the distance to a decision threshold. Within these models, affect could be posited to modify one or more of these and thereby explain individual differences in choice, such as, for example, incorrect responses that occur due to impulsivity (low decision-bound) vs. exaggerated uncertainty (low drift rate).

In what follows, we aim to give the reader an understanding of selected findings and theories in this literature, with a focus on models that offer mechanistic insights into the role of affect in mental illness. There are several ways that affect and bias towards certain affective states can be captured within existing modeling frameworks within the processes of learning and inference. Formalization using computational models could allow us to test these ideas directly, and so we review how each of these frameworks capture the development of affect and affect bias below.

*Reinforcement Learning Models*

Reinforcement learning (RL) refers to a class of machine learning algorithms that model how an agent takes actions, receives feedback in the form of rewards or punishment, and adjusts its expectations and behaviour to maximize long-term reward[23]. These models have long been used to study how individuals adapt their behaviour based on feedback from their environment[24] [25] [26]. Within the RL framework, affect may track specific patterns in the valence (reward or punishment) and magnitude of outcomes, which in turn affect learning and action selection. Here, the perceived magnitude of a rewarding outcome can be parameterized as a "reward sensitivity" parameter that scales its objective reward value. It has been shown that reduced reward sensitivity affects the learning process of individuals with depression[27] and that anhedonia is specifically related to reduced reward sensitivity (motivation to act on reward), in the absence of altered learning rates[28,29]. Dopamine agonism has also been shown to modify learning rates in healthy individuals[28].

Another line of work has related momentary happiness levels to the cumulative impact of recent outcomes and associated patterns of surprise. In one study using a simple a gambling task, for example, Rutledge et al. (2014) showed that trial-by-trial self-reported happiness was predominantly associated with the recent history of positive reward prediction errors (e.g. when outcomes were repeatedly better than expected) versus expected rewards themselves[30]. As prediction errors are weighted by learning rates in belief updating, biases in learning rates for positive versus negative outcomes may also bias individuals toward positive versus negative mood states (e.g., by effectively making positive surprises consistently stronger or weaker than negative surprises)[31].

In these models, one proposed role of affect is to bias reward sensitivity. Namely, in a more positive mood, rewards will be perceived as more rewarding; in a negative mood, punishments will feel more punishing. This can be adaptive in helping an organism track trajectories of change in the environment (e.g., seasonally changing resources), but it also creates the potential for instability. For example, if this biasing effect is too strong, a stable reward level can lead to exaggerated positive expectations, followed by disappointment and exaggerated negative expectations, followed by positive surprise and exaggerated positive expectations, and so on in cyclic fashion. This offers one interesting computational mechanism that could contribute to the emotional instability observed in bipolar or cyclothymic disorders[31]. Stronger effects of current mood on perceived outcomes have been observed in those with greater mood instability[32]. Current mood has also been shown to influence reward learning: Bennett et al. (2023) have demonstrated that positive affect induction leads to higher reward expectations, while negative affect induction leads to lower reward expectations[33]. Reward expectation was also driven by affect-related shifts in attention. That is, when in a positive affective state, individuals paid more attention to positive, high value features of the stimuli, while negative emotional states led individuals to pay more attention to negative, low-value features. In schizophrenia, negative symptoms have been associated with reduced reward-driven learning[34], reduced reward sensitivity[35], reduced learning on "Go" trials, as well as reduced tendency to explore potential actions (correlated with anhedonia)[36]. These results may tie together the early appearance of depressive symptoms as well as negative symptoms early in the psychosis prodrome[14]. However, it should be noted that the relationship between exploration and schizophrenia is complex. In particular, while reductions in *directed* exploration (i.e., directing choice toward more uncertain options) have been seen in some studies [37], other work has instead shown increased exploration (often via random choice strategies) in this population [38].

It is important to note that, while these RL models have included a role for positive or negative valence in action selection, there is a much wider array of known influences of affect and emotion not captured by this approach. Thus, it will be important to find ways in which these other influences (e.g., on attention, visceral regulation, interpretation biases, etc.) might be included in extended models (see [39,40]). In general, RL frameworks tend to model affect as the result of prediction errors related to rewards and punishment, with affect then shaping sensitivity to rewards.

*Active Inference*

Active inference is a Bayesian theoretical framework that models the brain as a predictive agent. In perception, active inference aims to identify interpretations of sensory input that maximize the predictive accuracy of an internal (generative) model of the world; in decision making, it aims to predict which actions are most likely to generate preferred (i.e., rewarding) outcomes[21]. The drive to reduce uncertainty through information-seeking actions (and thereby improve predictions) is a core feature of this framework. Due to its comprehensive nature, active inference has emerged as an effective framework for modelling affect, suggesting that it may arise, at least in part, from the brain's attempt to predict and explain both the world and internal bodily sensations[41] using an internal generative model. Dysfunctions within that generative model can lead to maladaptive affective responses in psychopathology, as the brain is unable to accurately predict the world or predict and regulate internal sensations[42].

A conceptual paper by Biddell *et al.* (2024)[43] outlined a theoretical framework based on active inference for how the interaction between affective arousal and perception of bodily signals may contribute to both emotional awareness and emotion regulation. Specifically, they propose that affect arises from an inference process where subjective experience and autonomic arousal jointly guide decision making toward effective means of reducing uncertainty. In this conceptualization, autonomic arousal tracks current uncertainty about one's environment[44]. *Arousal coherence* is then defined as the alignment of subjective experience and autonomic arousal. Building on active inference principles, the framework posits that the perception of autonomic arousal guides belief updating about affective states and leads to effective and adaptive regulation of those states [43]; that increased coherence is linked to well-being. One key limitation of this framework, however, is an assumption that the relationship between arousal coherence and emotion regulation is consistent, overlooking cases where decoupling of autonomic arousal and subjective experience might act as a temporary protective mechanism, such as in PTSD[45].

The above-mentioned framework also broadly aligns with work by Hesp *et al.* (2021)[46],[46] who developed a hierarchical active inference model to demonstrate how a simulated agent might infer the valence of its own affective states. A key piece of evidence in this *affective* inference process in their model is an internal observation of changes in confidence regarding one's own action model (i.e., whether predicted action outcomes are likely to be accurate). A key suggestion here was that increases and decreases in confidence act as evidence for positively and negatively valenced states, respectively, and that these affective states can then act as prior beliefs about future confidence levels. This link between affect and changes in confidence bears some similarity to the proposals (and evidence) mentioned above within RL that happiness levels go up and down based on positive and negative prediction errors. Using this model, they simulated a synthetic rat navigating a T-maze with positive and negative reinforcement to show the resulting behavioral dynamics. They found that affective states could allow for more effective use of recent changes in action-outcome contingencies (e.g., quicker changes in behavior after a simulated shock), whereas simulated rats without affective states would engage in maladaptive behavior (e.g. experiencing more punishments before learning to change strategy). The behavioral dynamics are nuanced, however, as negative affect promotes random exploratory behavior in the absence of learned habits, but instead causes an agent to fall back on ingrained habits (i.e., prior expectations about one's own actions) when these are present. While unique in positing a role for affect in governing the degree of random exploration (and habitual behavior) this proposal is also limited as it does not incorporate arousal, physiological states, attention, and other cognitive processes associated with affect (most notably perhaps is links between affective states, attention, and modulation of learning rates; e.g., as incorporated in[47] and[48].

In a more comprehensive approach, Smith & Lane (2015)[49] proposed that embodied emotions (emotions involving perception of bodily states) are appraised, perceived, and regulated through an iterative and hierarchical system operating at different time-scales (explaining the rapid and slow components of emotional responses)[49]. In this conceptualization, representations of emotionally relevant stimuli could be triggered by either sensation or imagined stimuli. When emotionally relevant content is deemed to be present, the body's internal states can be adjusted while changes are made to cognitive processes (such as allocation of attention). As certain cognitive processes occur more slowly, they can then regulate the initial emotional response in a top-down manner. Importantly, they map their proposed hierarchical model onto a putative neural hierarchy, implicating key regions for interoception and emotion processing, such as the insula and anterior cingulate cortex. Building on this work, they conceptually explored individual differences in emotional experience as being the result of differences in trait emotional awareness (tEA)[50]. Here, tEA refers to a set of factors influencing the perception and understanding of emotions: abstractness (linking sensations to concepts), granularity (conceptualizing affective states using fine-grained emotion concepts), and self-other distinction (the capacity to differentiate the emotions of self and others). They argue that differences in tEA could explain why people experiencing similar affective triggers may have different experiences and reactions - a concept familiar to any practicing therapist.

Smith *et al.* (2019)[42,51] subsequently explored how active inference could be used to model emotion *conceptualization* (i.e., the process of assigning conceptual meaning to cognitive and bodily responses to emotional stimuli). In their formal simulations, an agent was specifically modeled as inferring and reporting its emotional states based on both internal (interoceptive[52], cognitive) and external (contextual) cues[42], where each emotion category was mapped to a specific combination of these cues. In this setup, when the model was presented with inconsistent inputs, its confidence in classifying the emotion category decreased. When there were unpredictable transitions between emotions, the model also showed challenges in maintaining stable emotion concepts, leading to uncertainty and instability in its emotional state[53] (as might be seen, for example, in borderline personality disorder). The model could also be initialized with prior expectations about its likely emotions. If the model started with biased initial beliefs (e.g., overrepresenting negative emotions), it took longer to recognize other emotions (a state similar to depression). Finally, attention allocation was crucial; if the model focused more on external cues (e.g., social context), it developed weaker awareness towards its internal cues. For related proposals on emotion-related conceptualization processes and their link to interoception and predictive processing, see[54,55].

Motivated by the different proposals described above, Smith et al. (2020) (see replication in[5]) applied a Bayesian perception model to behavioral and electrocardiogram data in patients with anxiety, depression, eating disorders, or substance abuse disorders. Relative to a healthy sample, all patient groups showed a reduced ability to infer the timing of their own heartbeats in a cardiac interoceptive task[56]. Here, the model suggested this was due to reduced confidence in the cardiac signal, and not due to altered prior expectations. This suggested a specific mechanism whereby patients might have difficulty using interoceptive information to regulate visceral states or infer their emotional meaning.

In other work, active inference has been used to model approach-avoidance conflict, with increased decision uncertainty and altered drives toward affective avoidance demonstrated in those with anxiety, depression, and substance use disorders [57,58]. In general, active inference approaches can be used to model affect as a state determined by internal and external observations, and in terms of the confidence an agent has in their own model of the world.

As evident in this brief review of example approaches, active inference (and related Bayesian models of cognition) may offer broader resources than current applications of (model-free) RL for capturing the multidimensional nature of affect and emotion. However, empirical tests directly comparing these approaches to capture emotional dynamics remain to be done[59]. This will be an important direction for future work, as will increasing the number of studies which examine longitudinal trajectories of computational parameters[60].

*The Hierarchical Gaussian Filter*

The HGF[22] is a Bayesian model with links to other Bayesian approaches, such as active inference. Like in active inference, it assumes that agents model the causes of their sensory input and try to predict the inputs they will receive. Its hierarchical nature allows for the modelling of different levels of beliefs about sensory input and uncertainty about this input. As an example of this, we will discuss the modelling of the conditioned hallucinations task. In this task, participants learn a Pavlovian association between a visual and an auditory stimulus during the presentation of varying levels of auditory noise. They can then hallucinate the presence of the auditory stimulus when the visual stimulus is presented, but no auditory stimulus is. These "conditioned hallucinations" are increased in people with auditory hallucinations and with psychosis proneness in the general population and track hallucination severity over time[61–63]. The HGF applied to behavioral data for this task has three levels of beliefs. The first tracks whether the participant believes that the auditory stimulus was present; the second tracks the association between the visual and auditory stimulus; and the third tracks the perceived *volatility* (instability) of the association between the auditory and visual stimuli. In addition to these levels, there is a parameter that tracks the relative weighting of expectations (known as prior beliefs, or *priors*) and sensory observations. In general in Bayesian models, when priors (or other sources of information) are more *precise* (e.g. more certain or held with greater confidence), they are given more weight in perceptual decisions. Using this modelling approach, hyper-precise priors have been shown to drive conditioned hallucinations, and have been directly associated with neuroimaging[61].

The HGF has also widely been used to model behavior in various reversal learning tasks. Studies repeatedly show that psychotic-like experiences and positive symptoms were associated with increased perception of environmental volatility across social and non-social domains. For example, in social learning tasks, where participants need to infer the intentions of a confederate or an avatar, higher environmental volatility was associated with paranoid ideation[64,65], and shown to be higher in patients with schizophrenia[66]. Similarly, when predicting non-social outcomes in the predictive inference task, it was associated with delusional ideation in non-clinical and with positive symptoms in a clinical sample[67]. Interestingly, it has been associated with specific types of affect[68–70]. For example, expected volatility in the environment (and prior expectations about how volatility will change over time) has been associated with subjective stress levels[44], and more anxious individuals have shown different patterns of neural responses in the insula linked to prediction uncertainty[71]. Related approaches have also shown that anxiety is linked to greater tendencies to infer categorical changes in the hidden context (latent hidden states) generating observations[72], following on a prior body of work suggesting learning rates are greater, and less flexible, in those with higher anxiety. Within the Bayesian inference framework, and evidenced by the update equations of the HGF, greater learning rates can arise from both greater expected volatility or underestimations of sensory or observational noise. Once accounting for both in one model, higher schizotypal traits still relate to higher uncertainty about the environmental volatility, while subclinical symptoms related to depression and anxiety, as well as negative symptoms of patients with schizophrenia, are associated with reduced learning about the reliability of sensory information[67]. Taken together, this suggests that while psychotic symptoms are linked with issues in building more abstract statistical models of the environment, supported by cortical areas, negative affect is linked to misestimation of outcome uncertainty leading to greater expectations that the environment will change unpredictably over time, due to an aberrant bottom-up signal. Theoretically, this is quite sensible, as, if greater negative affect is taken as a sign that one's current model is inaccurate (i.e., due to repeated prediction errors in a changing environment), it will be optimal to increase trust in new observations and update beliefs more quickly[73], allowing one's model to once again reflect an accurate representation of the current environment and reduce the associated negative affect. However, if this uncertainty cannot be reduced (perhaps due to the source of uncertainty being within the observer rather than in the environment, which can at least be sampled differently to reduce uncertainty), this may sustain negative affect. This will be crucial in our discussion of the psychosis prodrome below. As such, the HGF can be used to model affect as a function of (potentially inaccurate) perceived uncertainty about the environment.

*Drift-diffusion models*

Roberts and Hutcherson (2019) reviewed the literature testing affective influences on decision making mechanisms within Drift Diffusion models (DDMs)[74]. Four parameters are usually estimated in DDM paradigms. The first is drift rate, which represents the speed at which information accumulates towards a certain decision threshold. In one study, those with high anxiety were found to show a higher drift rate towards recognizing threatening words, even when the valence of the word was irrelevant to the task[75]. This suggests they treated the threatening stimuli as carrying a more reliable signal. The second core parameter is the decision threshold, which represents the point at which enough evidence has been collected to make a decision. A higher decision threshold can lead to more cautious (and slower) decisions by favoring accuracy over speed. [74]note the complexity of how affect might influence decision threshold. On the one hand, negative feedback could increase decision thresholds (i.e., making behavior more cautious on subsequent trials). However, other negative experiences, such as anger, might instead decrease decision threshold and make behavior more impulsive.The third parameter in DDMs is the initial bias, which starts evidence accumulation closer to one decision threshold than the other. Here, one suggestion is that simple affective priming effects could bias this starting point in favor of congruent affective responses[76]. The last parameter is non-decision time, which represents the time required for sensory/perceptual processing of the stimuli and action initiation (i.e., processes unrelated to the decision itself). In this case, arousal levels associated with different emotional states, such as anxiety[77], may affect sensory/perceptual processing time, modifying non-decision time[74]. There is also evidence that drift rates may have direct clinical correlates. For example, Pe *et al.* (2013) found that specific symptoms (rumination) were associated with greater drift rate for negative stimuli, potentially linked to negative attentional biases [78].

While simple to use, responsive to affective states, and mappable to neurobiological processes[79–81], it is important to highlight that standard DDM approaches are also limited in a number of ways. For example, they do not directly model affect; rather, parameter estimates from the model can be correlated with (or experimentally modified) by affect in different contexts. In addition, they are restricted to decision-making contexts (though they can be generalized beyond binary decisions)[82] and not designed to model a number of other cognitive or perceptual processes relevant to affect.

Having reviewed a number of example models and representative findings related to the role of affect in computational psychiatry, we will now consider progress, challenges, and future directions for this field.

*Advances, Challenges, and Future Directions in Computational Models of Affect*

Recent advances in computational psychiatry have significantly enhanced our understanding of affective processes in psychiatric disorders. We identify three key strengths in the current literature. First, affect and mood are increasingly recognized as contextual factors that dynamically alter both current and future expectations, directly impacting learning and decision-making processes[42,51,56,74]. This perspective represents a major step forward in linking affective states with cognitive and behavioral outcomes in a systematic and quantifiable manner. Second, computational models have effectively captured the interplay between internal experiences and reactions to environmental stimuli, highlighting how individuals may differentially process the same external events depending on their internal emotional states and tEA[43,49]. This interplay suggests a personalized dimension to emotion processing, with implications for conditions such as depression and schizophrenia, where interoceptive disturbances have been observed[56,83]. Third, the literature has begun to integrate both state- and trait-like dimensions of affect, providing a more nuanced understanding of short-term affective fluctuations versus stable individual differences in emotional experience[42,50].

Despite these promising advances, several challenges remain. Current models have yet to fully integrate social and contextual influences on affect-driven learning and belief updating, limiting their applicability to real-world phenomena. Social cognition—including trust, cooperation, and emotional contagion—remains underexplored, despite compelling theoretical work suggesting that social reinforcement and feedback significantly shape affective processing [42,50,51,84]. Computational models have largely focused on individual decision-making in isolated contexts, neglecting the interactive and interpersonal dimensions of affect regulation. Another key limitation is the oversimplification of affect as a unidimensional (positive-negative continuum) construct. In addition, most models and experiments to date do not capture the longer-term temporal fluctuations of affect, which are dynamically shaped by prior experiences, uncertainty, and environmental context. Active inference models have begun addressing this issue by considering affect as arising within a hierarchical inference process that updates over time, yet the interactions between affective states, interoception, and social context require further refinement and empirical validation. We argue that comparing reinforcement learning (RL) and active inference within a unified framework presents a promising direction for future research[85].

There is a significant opportunity to use models incorporating affect to explore sex differences in the expression and treatment of psychiatric symptoms[86,87]. The influence of changing hormone levels on cognition and affect throughout the female reproductive lifecycle has often been either ignored or perceived as an unwanted source of variance, even though hormones have significant neurobiological effects[88]. This has led to biased recruitment, and a lack of understanding of how hormonal fluctuations can influence affective symptoms, such as in the premenstrual dysphoric syndrome[89–91]. Computational modelling of the dynamic interplay between affect, symptoms and hormones in females can therefore not only importantly inform future treatment development studies but can also serve as an excellent use-case for developing tools that are sensitive enough to capture subtle changes in affect over long time periods.

Taking these strengths and limitations into consideration, one key area for further advancement is in understanding the role of affect in longitudinal modifications of individuals' models of the world. The seeds of this are present in the current literature. For example, in a reinforcement learning context, pathological negative affective states may promote biased reward expectations, which can in turn lead to choices that reflect these (maladaptive) expectations. However, we argue that the current use of affect is the proverbial tip of the iceberg, and that the role of affect in shaping perception and explanatory models - for example, the form and content of prior beliefs that may drive psychotic symptoms[2,61,92–94] - may be more nuanced. In the following section, we will turn to a conceptual model that illustrates a putative role for affect in the development of psychotic symptoms.

*Potential role of affect in the psychosis symptom development and the prodromal period*

The earliest phases of psychosis constitute an affectively charged period. Retrospective studies of prodromal symptoms observed that depressive symptoms often emerge first, followed by negative symptoms, and then by attenuated psychosis symptoms[95,14,95,96,97]. Delusional mood[98,99], a sense of enhanced significance in the environment that needs explanation, occurs often prior to the development of frank delusions, which themselves tend to be negatively affectively valenced[100,101]. Finally, both people with clinical psychosis and those at high risk for psychosis are known to have increased sensitivity to stressful events, suggesting a phenotype of affective vulnerability[101,102]. As such, from the earliest phases of the prodrome to frank psychosis, there is a prominent role of affective symptomatology. Despite this, computational accounts of psychosis development have not comprehensively addressed the role of affect in a longitudinal sense. Here, we will attempt to extend our recent model of psychotic symptom development[2] to include an affective dimension, more fully encompassing known phenomenology.

Our model posits that the development of psychosis begins with increased cortical noise due to neurobiological factors such as NMDA receptor dysfunction. This increased cortical noise leads to aberrant prediction error signalling and impaired capacity of the brain to communicate and incorporate new information effectively. This intrinsic source of noise provides the germ for delusional belief formation, which has previously been linked to aberrant prediction error signalling (see [2,103,104]). From this point, multiple routes towards psychotic symptoms may be possible.

One possibility is that this biologically generated noise leads the brain to overestimate noise within sensory signals generated by the environment. If so, sensory inputs would be treated as unreliable sources of information, leading to a relative up-weighting of prior beliefs and slow learning (e.g., one would be more likely to perceive what is expected). This increased weight on prior beliefs is consistent with results of multiple studies using the conditioned hallucination task[61–63]. When making decisions, this high expected noise would also reduce confidence in one's model's ability to generate desired outcomes, corresponding to negative affect as posited within the active inference framework (as described above). In other words, the agent would experience negative affect because it infers it is not able to improve its ability to predict and control sensory input through action.

An alternative possibility is that this biologically generated noise is mistakenly treated as signal (i.e., as though it is reliable). In this case, instead of overestimating environmental noise, it would instead be forced to explain this input by inferring the state of the world is constantly changing (i.e., it would form the belief that environmental contingencies are highly unstable/volatile). This inference that the world is ever-changing would entail a down-weighting of prior beliefs and fast learning. In other words, the model would begin to overfit to noisy biological signals and show unstable fluctuations in beliefs over time. This is consistent with other work suggesting overfitting of visual and somatosensory/proprioceptive sensory input in psychosis (e.g., alterations in visual pursuit and force-matching, among others; see[105]). Even in this case, however, over longer timescales one might expect higher levels of processing to infer these constantly shifting beliefs are not effective at explaining sensory data and that this data may in fact be less reliable than initially estimated. This is speculative, but could represent a more indirect route to eventually down-weighting sensory signals and up-weighting prior beliefs.

Thus, one might assume that under both possibilities the longer-term result is a reduced reliance on sensory signals and a relative increase in reliance on prior beliefs. This compensatory increase in dependence on prior beliefs would then set the stage for frank psychosis. Specifically, psychosis onset could represent the point at which over-weighted (hyper-precise) priors finally overwhelm the corrective influence of sensory information, which had, until then, prevented these priors from generating threshold-level percepts[92,93]. Under this model, the attenuated psychotic symptoms seen in the prodrome would simply be caused by a less severe version of the imbalance between priors and sensations present during frank psychosis.

While this model does explain many of the observations in psychosis phenomenology (see[106] and [2] for discussion), it does have limitations. First, it does not address the early development of negative symptoms, of which affective symptoms are an integral component. Second, it does not address the cause of the valenced nature of delusions and hallucinations. To address these gaps, we add an affective component to the model below.

*Extension to cortical noise and negative symptoms*

First, we link the early increase in cortical noise to the development of negative symptoms, which have recently been shown to correlate with computational measures of low-level noise[67]. Cortical noise and the degeneration of grey matter [107] and white matter tracts [108,109] would lead to degradation in communication *between* cortical regions, in addition to bottom-up signal degradation that we have already tied to aberrant prediction error signaling. This can account for the distributed changes in multiple brain networks seen in negative symptoms [110]. This neural dysfunction has been shown to affect areas relevant to reward processing[111], consistent with an RL approach to understanding negative symptoms in terms of reduced reward sensitivity and reward-driven learning[34,35]. As such, anhedonia and avolition in negative symptoms can be conceptualized as a reduction in the integrity of reward processing. We discuss reduced expression of emotions in the section on social withdrawal. The proposed dysfunction in communication between brain regions also has important implications for affect dysregulation, as we argue in the following section.

*Noise and negative affect*

We propose that neural signal degradation, which initially results in the formation of negative symptoms, also plays a role in the development of a negatively valenced state during the prodrome. This likely underlies the development of negatively valenced hallucinations and delusions and would be consistent with the dysconnection hypothesis of schizophrenia[112], which posits large scale dysconnectivity between frontal and temporal regions underlying the illness. We argue that this dysconnectivity could lead to negative affect via dysfunction of emotion-related regions[113], such as the insula and amygdala, which also underlie key psychosis symptoms, such as paranoia and hallucinations[61,114,115]. In addition, increased perception of environmental instability (parametrized as volatility) during delusion formation can be pointed to as a potential source of negative (e.g. fearful, uneasy) affect[116], which could then be incorporated into the delusional belief.

*Negative belief-emotion spiral*

Consistent with the computational models described above, this negative affective state may cause the observed increase in sensitivity to stresses experienced in daily life seen in the high risk state[117] by continually biasing future expectations. Increased stress sensitivity creates a negative filter on information processing that may further reinforce negative beliefs and privilege negative observations. In an active inference framework, we might model this stress sensitivity as an increased precision (or learning rate) on negatively valenced observations. This would be present especially in the context of overall reduced sensory precision (i.e., greater expected noise in environmental signals)[62] or increases in expected environmental instability (volatility) early in the prodrome. This would then lead to an uncertain, fearful, and uneasy affective state that is potentially discordant with actual signals from the environment–in other words, delusional mood.

This interplay between affect and unusual beliefs is bi-directional, and it will extend through time. Negative beliefs will reinforce the negative affective state, and negative affect will reinforce negative beliefs. The observations present in the environment that might normally rescue this negative belief-emotion spiral (such as positive events or pleasant interactions with friends and family) are then limited in their ability to do so, as they are experienced with reduced precision, and may be sampled less extensively due to potentially reduced levels of directed exploration (i.e., selecting more uncertain options) caused by negative symptoms. This is in keeping with greater confidence in (likely negatively valenced) prior beliefs and consequent discounting of mood-incongruent sensory information. In fact, it has been observed empirically that individuals with psychosis show reduced levels of directed exploration[37], but increased levels of random exploration[38]. This is consistent with some other studies (but not all) suggesting greater negative affect is linked to lower directed exploration (reviewed in[118]). It is also notable that the active inference framework discussed above specifically predicts both lower directed exploration (i.e., less dependence on expected free energy) and greater random exploration (more noisy, value-independent choice) in those with greater negative affect.

*The Development of Valenced Positive Symptoms and Social Withdrawal*

Psychotic beliefs formed in the context of negative affect, and the privileging of negatively valenced information and internal states, can jointly account for the specifically negative valence of hallucinations and delusions in clinical populations. This belief formation occurs in the context of cortical signal degeneration (see[2]) and with a reduction in cognitive flexibility[119,120]. The longer this process persists, the stronger these priors become, and the less viable the alternative models of the world - resulting in the development of (negatively valenced) crystallized false beliefs[93,121]. This would explain why longer durations of untreated psychosis and more insidious symptom onset is associated with worsened outcome in schizophrenia [95,122,123].

Interestingly, reduced neural pathology in healthy voice hearers compared to patients with schizophrenia has been observed[124]. Under our model, this should result in less of the signal degradation and uncertainty that drives negative affect. This would then result in less negatively valenced hallucinations - a finding consistently observed in this population[125].

One additional symptom that may be better explained by this updated model is social withdrawal. An initial sense of fearfulness or uncertainty might increase aversion towards social engagement[126–128]. This would be consistent with the RL findings discussed above, which showed aberrant exploration under uncertainty in negative symptoms[36], and with other work linking alterations in exploration with negative affect and psychopathology more broadly[118,129–132]. In addition, consistent with RL modelling of negative symptoms, social anhedonia may cause increased social isolation[128], potentially as a result of reductions in reward sensitivity[35,133]. In line with the active inference frameworks described above, the person experiencing these symptoms would additionally observe themselves being less social, and enter self-states of loneliness, depression, social defeat, or social exclusion, further disincentivizing social interaction. While speculative, this negatively-biased conceptualization of their social situation could underlie reduced emotional expression in negative symptoms, which has previously been linked to reduced amplification of felt emotions in schizophrenia (despite similar felt emotions)[134]; e.g., if I believe that I am isolated from others, or that they may be a risk to me, and I have a reduced perception of the value of social interaction, I may be more likely not to place effort into making my emotions explicit (given that the goal of this is to facilitate interpersonal interaction). Recent work has begun to link altered learning under uncertainty in social contexts to negative symptoms, reinforcing the idea that negative symptoms result at least in part from an impaired understanding of social situations[64].

The self-reinforcing cycle of isolation would likely further degrade the amount and quality of contradictory information in the environment, reducing the stimuli which could counteract the growing prior hyper-precision[121]. Indeed, this reduction in directed exploratory behavior with negative affect is precisely what is suggested by the active inference approaches discussed above (e.g. [46]). These continuously increasing maladaptive prior beliefs would likely incorporate these observations as they are formed. This would explain the themes of self-deprecation, isolation from, and potentially of persecution by, others that are common in delusions and hallucinations.

It should be noted, however, that positive and negative symptoms do not necessarily co-occur, with positive symptoms tending to fluctuate more than negative symptoms, and that different computational correlates of negative vs. positive symptoms have been observed[67], which may call into question a role for negative symptoms and their associated low-level noise in the generation of positive symptoms. While this requires further longitudinal study, we suggest that this may be related to a separation in time. The low-level noise that generates negative symptoms is likely present prior to and regardless of whether positive symptoms are generated, but this noise provides the conditions in which negatively valenced priors can form, with these priors then driving hallucinations and delusional conviction independently of negative symptoms and low-level noise. Vice versa, negatively valenced priors may well form in other ways which are not associated with the presence of negative symptoms; longitudinal studies of symptom development with computational phenotyping would be required in order to identify these groups.

*Formalizing the addition of affect*

Affect would be added to this model formally in two components: a trait-like component, which would be roughly equivalent to the clinical definition of *mood*, and a state-like component, representing sensitivity to environmental cues and momentary affect (as discussed in the sections above). Each component could be further resolved into emotion networks composed of different positive and negative emotions, which would likely have differential activations dependent on factors such as premorbid personality and traumatic experiences. This would help account for the fact that delusions and hallucinations often contain content that is emotionally relevant to the individual's premorbid context, such as trauma[135,136]. In addition, it would be expected that the precise content of the eventual psychotic phenomena would depend on the precise affective context present during development. For example, sad and self-critical states during the prodrome would be expected to give rise to critical and self-deprecating hallucinations; and more fearful or anxious affect might be expected to give rise to more paranoid beliefs and hallucinations. In addition, previous work[137] has suggested that finding meaning through the construction of delusional narratives ("apophany" in classic phenomenology[99], and coincident with prior hyper-precision in our model), in the context of preceding states of uncertainty and lack of trust in the world, can be associated with more positive emotions.

We would expect that parameters representing these components would change in value over time, with negative affective states strengthening across the prodromal period. Additionally, we would expect reciprocal reinforcement between the trait- and state-like components, due to the negative belief-emotion spiral, making both relevant from a treatment planning perspective. In addition, we might expect to be able to decompose the state-like component to include an interoceptive dimension, consistent with both the active inference literature described above connecting interoceptive deficits with psychopathology and previous demonstration of somatic sensation alterations in schizophrenia[138,139]. Consistent with the Smith et al models described above, deficits in interoceptive inference may reduce the ability of patients developing psychosis to interpret, understand, and regulate their own emotions- making it both more difficult to regulate dysregulated affect, and increasing the uncertainty about the source of negative affect, which may further increase the need for highly precise, negatively valenced priors to "explain" the presence of the negative affect.

Modelling these affective components separately may be relevant for treatment planning, as temporary alleviation of affective state may not be effective as a treatment if a 'reservoir' of negative affectivity in the trait-like component (perhaps driven by maladaptively highly-precise priors which are difficult to update with low-precision observations[62]) will simply re-establish the negative state-like component in short order.

While this model would explain several phenomenological observations, in order to establish its validity it would be helpful to have specific predictions. Several of these are included below. Crucially, if validated through longitudinal computational phenotyping studies which can assess the temporal order of events with respect to changes in information processing, this model would provide a staging framework for the psychosis prodrome. Stages and parameter estimates at each stage (e.g., the strength of an individual's current prior beliefs) could be correlated with neurobiological signatures, and these neurobiologically-validated computational measures could then help identify novel treatment avenues and serve as biomarkers of illness progression and eventually as predictors of treatment effectiveness.

One area where this computationally-informed model of affect could have immediate and practical utility would be in better differentiating patients truly at risk for psychosis from those with attenuated psychotic symptoms in the context of trauma who will likely not develop a primary psychotic illness, given the high rate of trauma-related disorders in clinical high-risk samples[140]. Another area where this model may prove useful is in understanding sex differences in psychosis development. Women tend to experience more severe affective symptoms in psychosis as well as during psychosis risk[141,142]. In addition, men experience more negative symptoms, including increased asociality[97], and there is some evidence of differential performance in social appraisal in males versus females[142,143]. Computationally phenotyping men and women using tasks or approaches consistent with our proposed model may help clarify the causes of these sex differences. One area for further expansion of this model, while out of scope of the current work, is language, which is affected in psychosis and is core to both positive and negative symptoms[144]. Another key area will be to examine the influence of inter-individual variation in the development of psychotic themes. For example, many patients have grandiose delusions, rather than persecutory and negative delusions, and this model would need to be expanded in order to fully characterize how these more 'positive' positive symptoms could be explained. One possibility is that grandiose ideation is a *result* of negative affect- the "manic defense"[145]; another possibility is that those with grandiosity or mania have a more greater sensitivity to reward[146] (see[147] for discussion). Indeed both may be true in different people- highlighting that learning about subgroups of patients who deviate from the model we propose based on their premorbid cognitive or affective makeup will be key. One final area for further research is relationship of each element of this model with neurotransmitter signalling changes across the prodrome- for example, considering the finding that there is dopamine dysregulation prior to the onset of psychosis (see[148]).

*Model predictions*

1. If priors driving hallucinations/delusions are affectively laden, we would expect that tasks that tap into affective domains would produce stronger experiences when there is an affective compared to a neutrally valenced condition. We are currently investigating this with an affective version of the conditioned hallucinations task[61].
2. We should expect affective changes to precede the development of hallucinations and delusions in the psychosis prodrome if they are part of the causal chain that leads to the priors that drive them, and we should expect trauma to potentiate this process. As discussed, significant evidence exists for this prediction (e.g.[14,96]).
3. We would expect that the precision afforded to negatively valenced information would begin to increase early in the prodrome and would correlate with the strengthening of maladaptive priors (i.e., while the precision afforded to other information is also gradually decreasing). However, as noted, there may be inter-individual differences such that, for example, some prodromal individuals would potentially experience a bias towards positive valence, potentially leading to grandiosity.
4. We should expect there to be a temporal correlation between worsening affect and the onset of either increases in expected environmental instability (volatility) or increases in expected noise in sensory input (i.e., lower sensory precision-weighting, increased weighting of prior beliefs), as these are thought to potentially underlie negative affect.
5. We would expect that affective themes experienced earlier in prodrome (and influenced by premorbid experiences, trauma and cognitive and affective makeup) should be replicated in the priors driving hallucinations and delusions later on, and therefore in the themes of these symptoms. Here we would expect to be able to examine and parse inter-individual differences and define potential subgroups within our proposed model.
6. Treatment of depressive symptoms earlier in the prodrome should reduce the intensity of negative affect present in positive symptoms, especially for self-critical psychotic symptoms; this would be due to a modulation of the intensity of the affect during the generation of the maladaptive prior. However, as affect *shape* but does not *cause* prior hyper-precision, this would not necessarily prevent the psychosis from onsetting.
7. After controlling for prodrome length, we would expect measures of sensitivity to stress during the prodrome (such as performance on stressful tasks[149]) to correlate with increased symptom severity after illness onset.
8. Treating affective states early in the prodrome should change the intensity and content of symptoms; however, since affective change provides context to the developing symptoms, but is not the primary process, treating them would not necessarily prevent the development of psychosis.

CONCLUSION

In this review, we have discussed how computational approaches have enabled us to better understand the role of affect in information processing within psychiatric illness. Using these principles, we have extended a recent model of psychotic symptom development to include affect as an important source of context during symptom development. Given the fact that affective states can be modified, they may serve as an important focus of treatment during the prodrome. Likewise, considering and parsing different affective states during illness and illness development using computational tasks may help improve diagnosis, prognosis, and treatment planning in a range of psychiatric conditions.

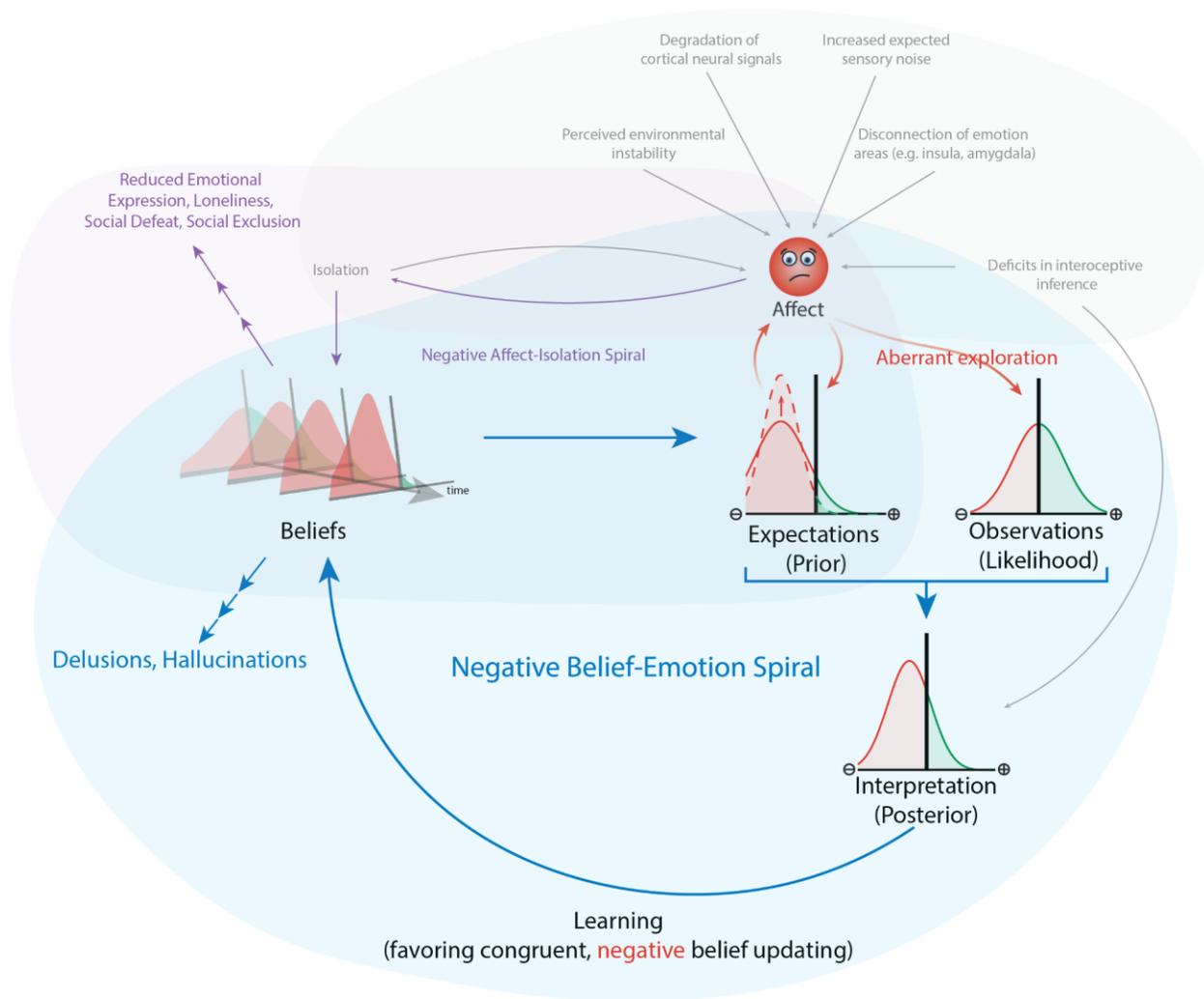

**Figure 1.** Negative Belief-Emotion Spiral (Blue): Within our Bayesian conceptualization of the prodrome, expectations (priors) form the lens by which incoming evidence (observations) is interpreted. Negative affect may increase the precision on negative observations and limit sampling of corrective, positive observations (via aberrant exploration). Negative interpretations may subsequently reinforce negative beliefs, further biasing future expectations and interpretations. Over time, beliefs crystallize, shaping negatively valenced delusions and hallucinations. Negative Affect-Isolation Spiral (Purple): Negative affect may also inhibit social engagement and underlie negative conceptualizations of one's sociability. This would result in outward symptoms of loneliness, social defeat, and even limited social expression. Biopsychosocial Underpinnings of Negative Affect (Gray): Negative affect likely stems from biological (degradation of cortical signals, disconnection of emotion areas), information processing (e.g. perceived environmental instability, greater expected noise, deficits in interoception) and social (e.g. isolation) factors.